# Hadamard single-pixel imaging versus Fourier single-pixel imaging

**Zibang Zhang,[1] Xueying Wang,[1] Guoan Zheng,[2] and Jingang Zhong[1,3,*]**

[1]*Department of Optoelectronic Engineering, Jinan University, Guangzhou 510632, China*
[2]*Biomedical Engineering, University of Connecticut, Storrs, CT, 06269, US*
[3]*Guangdong Provincial Key Laboratory of Optical Fiber Sensing and Communications, Jinan University, Guangzhou 510632, China*
*\*Corresponding author: tzjg@jnu.edu.cn*

**Abstract:** Single-pixel imaging is an innovative imaging scheme and has received increasing attentions in recent years. It is applicable to imaging at non-visible wavelengths and imaging under low light conditions. However, single-pixel imaging has once encountered problems of low reconstruction quality and long data-acquisition time. This situation has been changed thanks to the developments of Hadamard single-pixel imaging (HSI) and Fourier single-pixel imaging (FSI). Both techniques are able to achieve high-quality and efficient imaging, remarkably improving the applicability of single-pixel imaging scheme. In this paper, we compare the performances of HSI and FSI with theoretical analysis and experiments. The results show that FSI is more efficient than HSI while HSI is more noise-robust than FSI. Our work may provide a guideline for researchers to choose suitable single-pixel imaging technique for their applications.

## 1. Introduction

Contemporary single-pixel imaging originates from ghost imaging [1-10]. Ghost imaging was initially considered as a quantum effect [1] but later Bennink et. al. demonstrated that it can be implemented with a classical source [2]. Computational ghost imaging [3-10] allows one to capture a scene using a single-pixel (or bucket) detector. Objects to be imaged are under spatially and temporally varying illuminations. The illumination patterns are typically generated using a spatial light modulator (SLM). The single-pixel detector is used to collect the corresponding light signal for each illumination pattern. The desired image is computationally reconstructed by correlating the illumination patterns with the detected signals.

Single-pixel imaging techniques [11-29] do not need to use any pixelated detector for light signals detection. This advantage brings single-pixel imaging a potential capability of solving some challenges in conventional imaging. For example, single-pixel imaging scheme allows one to build a low-cost imaging system that can work at non-visible wavelengths. It should be noted that pixelated detectors are commonly expensive or even unavailable at most non-visible wavelengths. In comparison with pixelated detectors, single-pixel imaging may also have the advantage of imaging at low light conditions. This is because single-pixel detectors with large active area are easier to fabricate and more sensitive to light.

Inherited from ghost imaging, single-pixel imaging was initially based on a statistical model, which can be evidenced by the use of random patterns for illumination. The random patterns form an overcomplete non-orthogonal basis. Consequently, it requires a great number of measurements (much larger than pixel counts) and long data-acquisition time for recording signals. Even with so many measurements, the quality of images reconstructed by single-pixel imaging is hardly comparable with the quality of images obtained by conventional 2D-detector-based imaging. Long data acquisition time and low-quality reconstruction limit the applicability of this innovative imaging scheme. Such a situation existed till the emergence of single-pixel imaging techniques based on a deterministic model.

Hadamard single-pixel imaging (HSI) [12-22] and Fourier single-pixel imaging (FSI) [23-27] are two single-pixel imaging techniques that use a deterministic model. Both techniques employ deterministic basis patterns for illumination -- HSI uses Hadamard basis patterns for illumination while FSI uses Fourier basis patterns. The use of basis patterns for illumination brings two advantages. The first advantage is perfect image reconstruction. The basis patterns form a complete orthogonal set and the use of basis patterns for illumination allows one to acquire the spatial information of object image in a transformation domain. When is fully sampled in the transformation domain, the image can be losslessly reconstructed by the corresponding inverse transform. This feature solves the problem of low-quality reconstruction in ghost imaging. The second advantage is measurement reduction. Natural images give a sparse representation in either Hadamard or Fourier domain, allowing one to reconstruct a sharp image with under-sampled data. This feature solves the problem of long acquisition time. As such, HSI and FSI well tackle the problems that are inherited from ghost imaging.

In this paper, we theoretically and experimentally compare these two techniques, in terms of principles, imaging efficiency, noise robustness, and etc. This comparison shows the commons and the differences between HSI and FSI. The comparison also presents the advantages and disadvantages of the both techniques.

## 2. Comparison of theory

### 2.1 Principle of HSI and FSI

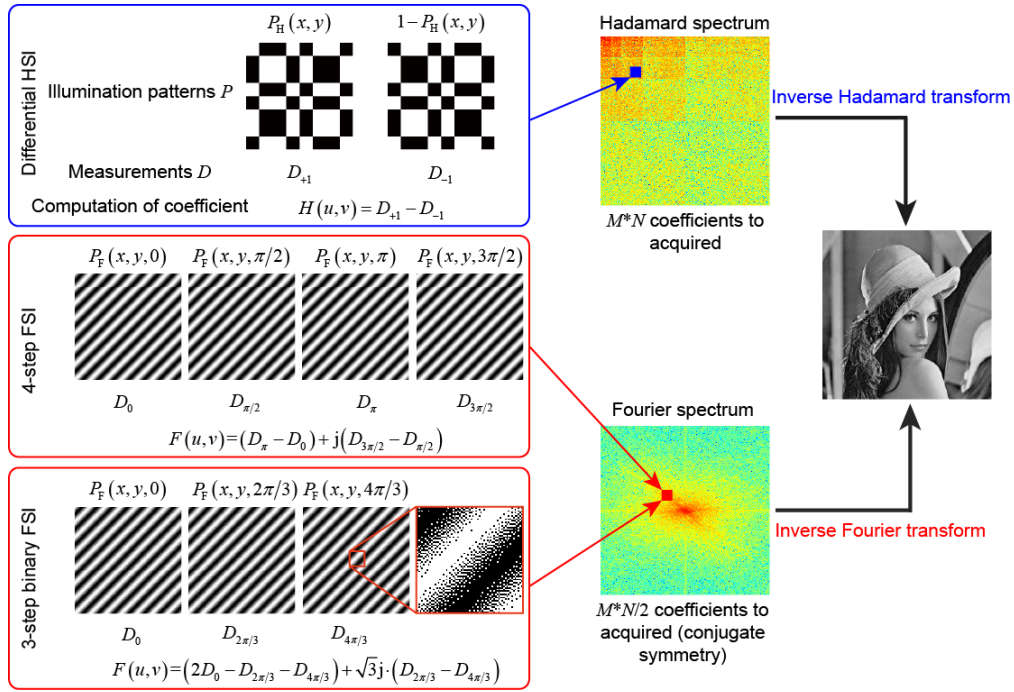

Fig. 1. Illustration of differential HSI, 4-step FSI, and 3-step binary FSI.

HSI is based on Hadamard transform [30]. HSI acquires the Hadamard spectrum of the object image and reconstructs the object image by applying an inverse Hadamard transform. Hadamard spectrum is composed by a group of Hadamard coefficients. Each coefficient corresponds to a unique Hadamard basis pattern. To obtain a Hadamard coefficient, one can project the corresponding Hadamard basis pattern(s) onto the object and use a single-pixel detector to measure the resultant light intensity. The single-pixel light intensity measurement is mathematically equivalent to the inner product between the Hadamard basis pattern(s) and the object. As such, the Hadamard spectrum can be reconstructed based on the single-pixel measurements. The two-dimensional Hadamard transform $H\{\ \}$ of an image $I(x,y)$ is defined as

$$\tilde{I}_\mathrm{H}(u,v) = H\{I(x,y)\} = \sum_{x=0}^{M-1}\sum_{y=0}^{N-1} I(x,y)(-1)^{q(x,y,u,v)}, \tag{1}$$

where

$$q(x,y,u,v) \equiv \sum_{i=0}^{n-1}\left[g_i(u)x_i + g_i(v)y_i\right] \tag{2}$$

and

$$\begin{aligned}
g_0(u) &\equiv u_{n-1} \\
g_1(u) &\equiv u_{n-1} + u_{n-2} \\
g_2(u) &\equiv u_{n-2} + u_{n-3} \\
&\vdots \\
g_{n-1}(u) &\equiv u_1 + u_0
\end{aligned} \tag{3}$$

where $n = \log_2 N$. Hadamard transform is only applicable for input images of size $N$-by-$N$, where $N$, $N/12$, or $N/20$ is a power of 2.

A Hadamard basis pattern $P_\mathrm{H}(x,y)$ can be obtained by applying an inversed Hadamard transform to a delta function $\delta_\mathrm{H}(u,v)$,

$$P_\mathrm{H}(x,y) = \frac{1}{2}\left[1 + H^{-1}\{\delta_\mathrm{H}(u,v)\}\right], \tag{4}$$

where $H^{-1}\{\ \}$ denotes an inverse Hadamard transform and

$$\delta_\mathrm{H}(u,v) = \begin{cases} 1, & u = u_0, v = v_0 \\ 0, & \text{otherwise} \end{cases}. \tag{5}$$

Differential HSI is an embodiment of HSI, allowing each Hadamard coefficient $H(u,v)$ to be acquired in a manner of differential measurement. Differential HSI is conducive to depression of noise. As illustrated in Figure 1, to acquire a coefficient $H(u,v)$, differential HSI takes two measurements. The one coefficient is acquired by projecting a Hadamard basis pattern $P_\mathrm{H}(x,y)$ and the other coefficient is by its inverse $[1 - P_\mathrm{H}(x,y)]$. The coefficient $H(u,v)$ is obtained by using the two corresponding measurements

$$H(u,v) = D_{+1} - D_{-1}, \tag{6}$$

where $D_{+1}$ and $D_{-1}$ are measurements corresponding to the illuminations of $P_\mathrm{H}(x,y)$ and $[1 - P_\mathrm{H}(x,y)]$, respectively. Hadamard coefficients are real-valued and the number of Hadamard coefficients is the same as that of image pixels. Fully sampling an $M \times N$-pixel image using differential HSI takes $2 \times M \times N$ measurements.

FSI is based on Fourier transform. FSI acquires the Fourier spectrum the object image and reconstructs the object image by applying an inverse Fourier transform. Fourier spectrum is composed by a group of Fourier coefficients. Each coefficient corresponds to a unique Fourier basis pattern. To obtain a Fourier coefficient, one can project the corresponding Fourier basis pattern(s) onto the object and use the single-pixel detector to measure the inner product between Fourier basis patterns and the object. As such, the Fourier spectrum can be reconstructed from the single-pixel measurements. Fourier basis patterns are also known as sinusoidal patterns or fringe patterns. Fourier transform was proposed in the late 1800s by Joseph Fourier and has been widely used in a number of fields. It allows any signal to be decomposed into a set of orthogonal sinusoidal waveforms of different frequencies. For example, images as two-dimensional signals are allowed to be broken down into a combination of sinusoidal intensity patterns. The two-dimensional Fourier transform $F\{\ \}$ of an image $I(x,y)$ is defined as

$$\tilde{I}_F(u,v) = F\{I(x,y)\} = \sum_{x=0}^{M-1}\sum_{y=0}^{N-1} I(x,y)\exp\left[-j2\pi\left(\frac{ux}{M}+\frac{vy}{N}\right)\right], \quad (7)$$

A Fourier basis pattern $P_F(x,y)$ can be obtained by applying an inverse Fourier transform to a delta function $\delta_F(u,v,\varphi)$,

$$P_F(x,y) = \frac{1}{2}\left[1+\left|F^{-1}\{\delta_F(u,v)\}\right|\right], \quad (8)$$

where $F^{-1}\{\ \}$ denotes an inverse Fourier transform and

$$\delta_F(u,v,\varphi) = \begin{cases} \exp(j\varphi), & u=u_0, v=v_0 \\ 0, & \text{otherwise} \end{cases}. \quad (9)$$

4-step FSI and 3-step FSI are two embodiments of FSI, using the 4-step phase-shifting formula (Eq. (10)) and the 3-step phase-shifting formula (Eq. (11)), respectively. Both embodiments allow each Fourier coefficient $F(u,v)$ to be acquired in a manner of differential measurements. 4-step FSI and 3-step FSI acquire each Fourier coefficient with 4 and 3 measurements, respectively. Please note that Fourier coefficients are complex-valued. 4-step FSI allows one to acquire each complex-valued Fourier coefficient $F(u,v)$ by projecting four patterns $P_F(x,y,0)$, $P_F(x,y,\pi/2)$, $P_F(x,y,\pi)$ and $P_F(x,y,3\pi/2)$, and using the corresponding four measurements $D_0$, $D_{\pi/2}$, $D_\pi$, and $D_{3\pi/2}$:

$$F(u,v) = (D_\pi - D_0) + j(D_{3\pi/2} - D_{\pi/2}). \quad (10)$$

Also note that $P_F(x,y,\pi)$ is the inverse of the pattern $P_F(x,y,0)$; $P_F(x,y,3\pi/2)$ is the inverse of the pattern $P_F(x,y,\pi/2)$. The number of Fourier coefficients is the same as that of image pixels ($M \times N$). With the prior knowledge that the Fourier spectrum of any real-valued image is conjugated symmetric, fully sampling an $M \times N$-pixel image using 4-step FSI takes $2 \times M \times N$ measurements. It can be seen that the 4-step FSI is essentially a differential method of measurement. FSI can also be conducted by employing the 3-step phase-shifting formula. The 3-step FSI acquires each complex-valued Fourier coefficient with 3 measurements,

$$F(u,v) = (2D_0 - D_{2\pi/3} - D_{4\pi/3}) + \sqrt{3}j\cdot(D_{2\pi/3} - D_{4\pi/3}), \quad (11)$$

where $D_0$, $D_{2\pi/3}$, and $D_{4\pi/3}$ are the measurements corresponding to the illumination patterns of $P_F(x,y,0)$, $P_F(x,y,2\pi/3)$, and $P_F(x,y,4\pi/3)$, respectively. Evidenced by Eq. (11), 3-step FSI is also a differential method of measurement, but in an asymmetric form. As the 4-step FSI does, the 3-step FSI has the property of noise suppression, but its performance is not as good as the 4-step FSI. It is because the 4-step FSI is a differential method of measurement in a symmetric manner while the 3-step is in an asymmetric manner. Fully sampling an $M \times N$- pixel image using 3-step FSI takes $1.5 \times M \times N$ measurements. In comparison with 4-step FSI, 3-step reduces 25 % measurements.

FSI and HSI are single-pixel imaging techniques based on basis scan and theoretically allow perfect reconstruction for any images in noiseless situations.

*2.2 Basis patterns generation*

The core of single-pixel imaging is to employ active illumination to acquire the spatial information of a target object. Instead of using random patterns, basis-scanning single-pixel imaging techniques use deterministic basis patterns for illumination. Figure 1 shows the comparison between the Hadamard and Fourier basis patterns. The difference can be summarized as follows: 1) Hadamard basis patterns are binary and mosaics look-alike while Fourier basis patterns are grayscale and fringes look-alike; 2) Hadamard basis patterns only have horizontal and vertical features while Fourier basis patterns have horizontal, vertical, and oblique features; 3) Fourier basis patterns are strictly periodical while Hadamard basis is not.

The applicability and the performance of these basis scanning single-pixel imaging techniques rely on the effectiveness and efficiency of basis patterns generation. Thus, it is necessary to discuss the methods of basis patterns generation for HSI and FSI. Hadamard basis patterns are binary (black-and-white), which makes HSI naturally suitable for single-pixel imaging systems based on a digital micro-mirror device (DMD). As DMD is a binary device, HSI can benefit from the high-speed binary illumination ability given by a DMD. Additionally, binary Hadamard basis patterns would not lead to quantization errors or gamma (nonlinear) distortion. However, to our best knowledge, Hadamard transform does not carry clear physical meaning, which makes Hadamard basis patterns almost impossible to be generated by a physical means. The applicability of HSI relies on the use of SLM. In other words, HSI would likely have difficulties when SLMs are unavailable at certain spectral regions.

On the other hand, Fourier basis patterns are naturally grayscale. It leads to the fact that FSI is not able to take the benefit of a high-speed SLM, such as DMD. DMDs work much slower in the grayscale mode. For example, a state-of-art DMD can display ~20,000 binary patterns per second but can only display ~250 8-bit grayscale patterns per second. Thus, FSI suffers from slower illumination rate and therefore longer data-acquisition time than HSI. In addition, using any digital device to generate Fourier basis patterns would cause quantization errors. When the number of quantization levels is too low, the resulting quantization errors would lead to pronounced image quality degradation. Recently, Zhang *et al.* proposed binary FSI [25], a workaround that uses binary Fourier basis patterns for illumination. Binary Fourier basis patterns are generated by upsampling and then dithering the grayscale Fourier basis patterns. The shortcoming of this approach is at the expense of reduced spatial resolution. Fortunately, Fourier transform is a natural operator and has physical meaning, which enables Fourier basis pattern to be generated by some physical means. For example, an ideal thin lens can be used as a Fourier transform engine with which the Fourier transform of an object image can be obtained at the back focal plane of a thin lens. Even without a thin lens, the Fourier transform of an object image can be approximately obtained by the far-field diffraction pattern, which is subject to Fraunhofer diffraction. Thus, one can generate Fourier basis patterns by using the interference of two plane waves, which adds applicability to FSI, especially for the cases that SLMs are expensive or even not available.

In short, FSI is more flexible and variant than HSI in terms of illumination patterns generation while HSI can benefit much more from the high-speed binary DMDs.

*2.3 Robustness to noise*

Both HSI and FSI are robust to dark noise and read-out noise in principle, because, as evident by Eq. (1) and Eq. (7), Hadamard transform and Fourier transform are global transformation. Global transformation has a property that each point (coefficient) in the transformation domain is a weighted sum of all points in the spatial domain and each point (image pixel) in the spatial domain is also a weighted sum of all points in the transformation domain. In other words, global transformation implies that each pixel in a reconstructed image is contributed by all measurements and such an averaging process allows evening out errors in measurement.

In terms of quantization errors, HSI outperforms FSI. It is because Fourier basis patterns continuously vary in space and magnitude. Thus, quantization errors will be caused when using digital devices (such as DMD or Liquid Crystal on Silicon (LCoS)) to generate Fourier basis patterns. Such errors would degenerate the quality of final reconstruction. On the other hand, Hadamard basis patterns are naturally in a discrete manner. Digital devices are able to generate error-free Hadamard basis patterns.

*2.4 Efficiency*

We refer efficient single-pixel imaging to a technique that allows one to reconstruct a sharp image with a small number of measurements. Additionally, highly efficient single-pixel imaging enables time-lapse imaging. Since the throughput of a single-pixel imaging system is inherently limited by the readout rate of the single-pixel detector, it is of critical importance to improving the efficiency of single-pixel imaging techniques.

The efficiency of a basis scanning single-pixel imaging technique depends on how well the utilized transformation concentrates the image energy. If a transformation is able to highly concentrate the image energy within a waveband (i.e., a small number of coefficients have large magnitudes), one can simply measure those large-magnitude coefficients and omit the rest to improve efficiency.

For natural images, energy is usually uniformly distributed in the spatial domain. Both Hadamard transform and Fourier transform have the ability to concentrate the image energy near the origin of their transformation domain. Thus, both HSI and FSI are able to reconstruct an imaging under Nyquist conditions by sampling coefficients with large magnitude. As will be quantitatively demonstrated in the following experimental comparison section, for natural images, Fourier transform gives a more condensed representation than Hadamard transform does. In other words, FSI outperforms HSI in terms of energy concentration.

Moreover, the physical meaning of Fourier transform gives it the capability of characterizing the impulse-response of some optical system. Optical transfer function (OTF), defined as the Fourier transform of the point spread function, shows how different spatial frequencies are handled by the system. For most optical systems, the OTF is an equivalent low-pass spatial filter. As a result, the spatial information that an imaging system can collect is low-pass filtered.

It can be seen that the characteristics of natural images and optical imaging system concentrate the image energy at the lower frequencies range of the Fourier domain. As such, it leads to rapid convergence in terms of reconstruction quality for FSI to sample along the increment of spatial frequency.

In some diffraction-limited situations such as microscopy, the achievable spatial resolution is determined by the cut-off spatial frequency of the optical system. In this case, the optical system acts as a low pass filter with a finite ability to resolve details. In mathematical terms, a diffracted-limited system sets a spatial cutoff frequency $f_0$ in the Fourier domain. Only the spatial information that falls into the circular low-frequency band with a radius of $f_0$ can be acquired through the system. Thus, FSI using a circular sampling path is the most effective approach for a diffraction-limited system in principle. If the spatial cutoff frequency of the system is given, FSI allows one to sample the circular low-frequency band only. Even if the spatial cutoff frequency is not given, FSI allows one to sample the Fourier domain along a circular path until the sampled energy is lower than a certain threshold.

Therefore, lower spatial frequency components are typically of much larger magnitude than higher spatial frequency components, as shown in Fig. 1. This prior knowledge implies a simple but efficient sampling strategy that only low-frequency components are needed to be sampled in the data-acquisition process.

## 3. Comparison of experiment

### 3.1 Numerical simulations

We firstly compare the performance of both techniques through numerical simulations using MATLAB. In the presented simulations, we focus on the quality of reconstructed images by using both techniques, especially when the images are undersampled. It should be emphasized that the throughput of single-pixel imaging systems is, in general, lower than conventional imaging systems and undersampling is a straightforward approach to reduce the number of measurements. Undersampling refers to acquire only the low-frequency coefficients and omit the high-frequency ones, given the prior knowledge that most energy of natural images is concentrated in low-frequency bands. Since sampling strategies may change the quality of the reconstructed images, we use three different sampling strategies for HSI and FSI without losing generality. The sampling strategies refer to the path along which coefficients are to be acquired. The sampling strategies in comparison are shown in Fig. 2. We note that the origin for Fourier spectra is located at the center of the pictures while the origin for Hadamard spectra is located at the left-top corner of the pictures. Therefore, Fourier spiral path is equivalent to Hadamard square path; Fourier diamond path is equivalent to Hadamard square path; Fourier circular path is equivalent to Hadamard circular path.

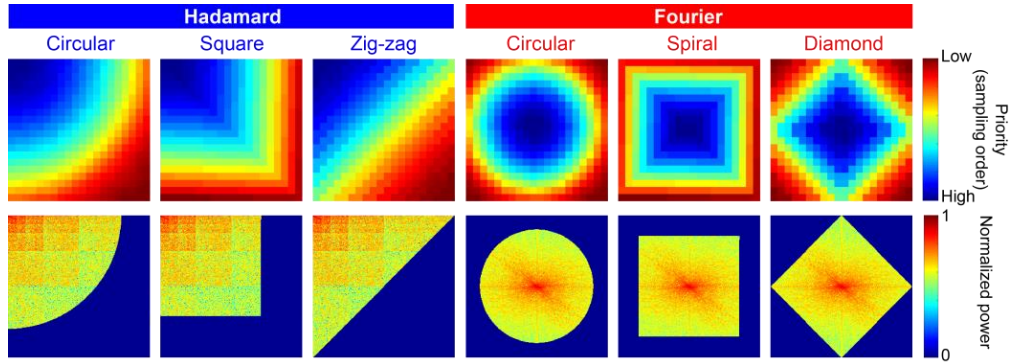

Fig. 2. Illustration of six sampling strategies used in our simulations. The pictures in the first row are sampling paths, according to which coefficients are acquired from low-frequency bands (blue) to high-frequency bands (red). The pictures in the second row are sample spectra acquired along the corresponding path. Spectra shown are with the sampling ratio of 50%.

Three different and characterized input images are used in the comparison, the 1951 USAF test chart pattern, 'Siemens star' target pattern, and the Lena image. The 1951 USAF test chart pattern is a binary pattern consisting of groups of three bars with dimensions from big to small. The bars are along vertical and horizontal directions. The 'Siemens star' target pattern is also a binary pattern. It provides multiple contrast measurements from a wide range of spatial frequencies. Different from the USAF 1951 test chart pattern, the 'Siemens star' target pattern has many oblique features. The Lena image is one of the most widely used natural image in imaging system tests and has multiple gray levels. We use peak signal-to-noise ratio (PSNR), structural similarity index (SSIM), and power ratio to quantitatively evaluate the quality of reconstructed images. Power ratio is a quantity which can evaluate the energy concentration ability, defined as power of the acquired spectrum to that of the complete spectrum ratio. Please note that, for Tables 1-3, the data in blue are the best results for HSI, the data in red are the best results for FSI, and the data in bold are the overall best results.

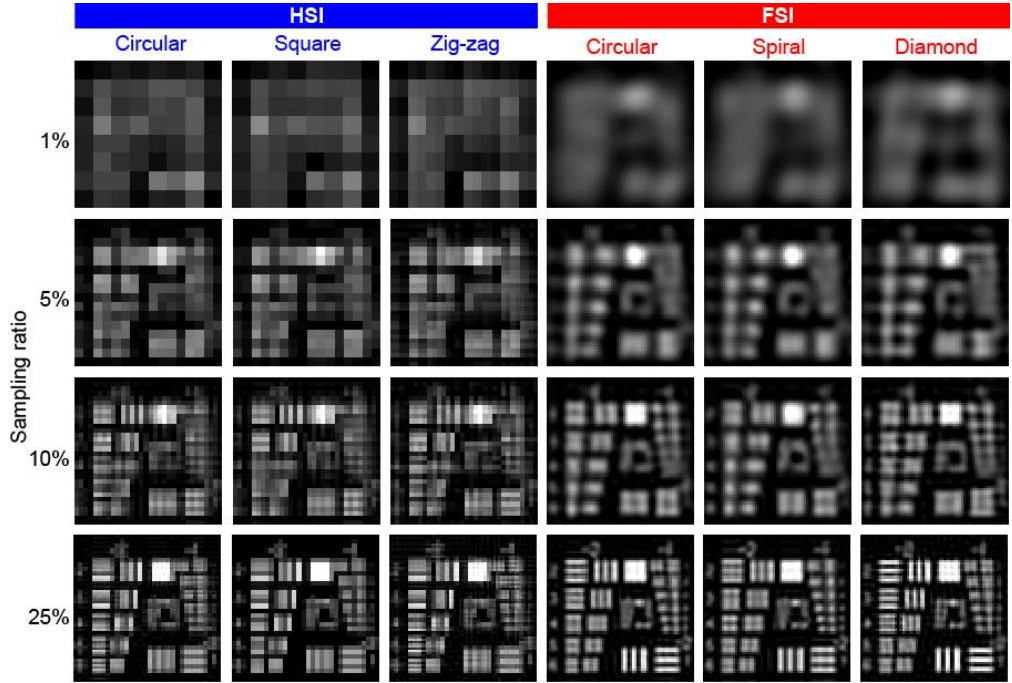

Fig. 3. Comparison results for USAF 1951 test chart pattern reconstruction by HSI and FSI for different sampling ratios.

Table 1. Quantitative comparison results for USAF 1951 test chart

|  |  | Strategy | Sampling ratio | | | | | | | |
|---|---|---|---|---|---|---|---|---|---|---|
|  |  |  | 1% | 5% | 10% | 15% | 20% | 40% | 60% | 80% |
| PNSR (dB) | Hadamard | circular | 11.00 | 12.45 | 13.17 | 13.64 | 14.25 | 15.45 | 19.42 | 27.24 |
|  |  | square | 10.93 | 12.39 | 13.04 | 13.49 | 13.84 | 15.09 | 16.68 | 20.94 |
|  |  | zig-zag | 11.10 | 12.47 | 13.34 | 13.89 | 14.42 | 17.62 | 22.85 | 27.94 |
|  | Fourier | circular | 11.39 | 12.71 | 13.55 | 14.22 | 15.52 | 19.81 | 24.27 | **33.01** |
|  |  | spiral | 11.37 | 12.66 | 13.28 | 13.91 | 14.51 | 18.48 | 22.19 | 26.59 |
|  |  | diamond | **11.47** | **12.77** | **13.72** | **15.01** | **16.27** | **21.58** | **26.71** | 33.00 |
| SSIM (%) | Hadamard | circular | 7.2 | 30.0 | 41.1 | 48.3 | 57.2 | 67.2 | 83.1 | 93.2 |
|  |  | square | 9.3 | 28.4 | 39.7 | 48.5 | 51.4 | 69.0 | 74.8 | 86.4 |
|  |  | zig-zag | 9.1 | 28.2 | 41.6 | 48.5 | 57.2 | 71.4 | 88.1 | 95.9 |
|  | Fourier | circular | 9.5 | 31.0 | **49.2** | 59.0 | 67.4 | 87.1 | 93.5 | 98.4 |
|  |  | spiral | 9.9 | 31.8 | 44.3 | 54.8 | 62.7 | 84.1 | 91.6 | 95.9 |
|  |  | diamond | 9.2 | **32.0** | **49.2** | **59.8** | **70.0** | **86.5** | **94.4** | **98.8** |
| Power (%) | Hadamard | circular | 6.7 | 18.5 | 27.5 | 33.7 | 40.1 | 56.6 | 76.0 | 91.8 |
|  |  | square | 6.4 | 18.0 | 26.5 | 32.6 | 37.5 | 53.7 | 69.0 | 86.1 |
|  |  | zig-zag | 7.3 | 18.6 | 28.4 | 35.2 | 41.2 | 63.2 | 81.4 | 92.1 |
|  | Fourier | circular | 7.7 | 20.0 | 29.2 | 36.6 | 45.1 | 69.0 | 84.2 | 95.1 |
|  |  | spiral | 7.6 | 19.8 | 27.5 | 34.6 | 41.0 | 65.6 | 80.9 | 91.5 |
|  |  | diamond | **8.3** | **20.7** | **30.6** | **40.4** | **48.7** | **73.1** | **86.9** | **95.2** |

As the results shown in Fig. 3 and Table 1, FSI with the diamond path gives the best reconstruction for the USAF 1951 test chart pattern. With the diamond path, Fourier coefficients on both spatial frequency axes are acquired preferentially. Those Fourier coefficients corresponding to horizontal and vertical fringe patterns highly correlate with the features of the resolution target (that is, horizontal and vertical bars). Thus, FSI with diamond sampling strategy achieves better reconstruction than the others for the same sampling ratio.

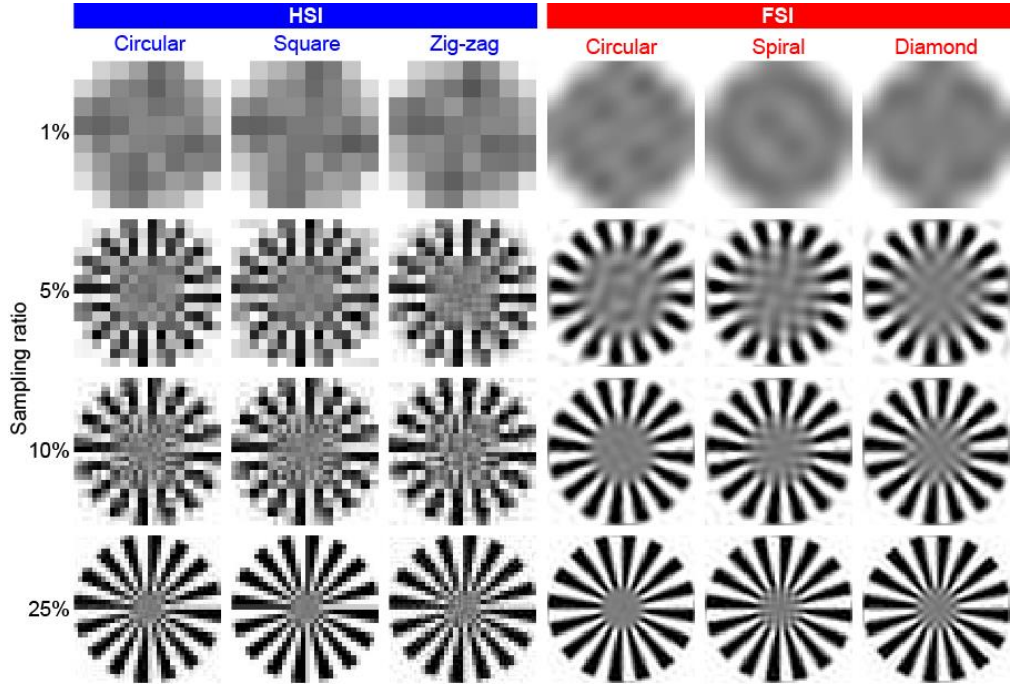

Fig. 4. Comparison results for "Siemens star" target pattern reconstruction by HSI and FSI for different sampling ratios.

Table 2. Quantitative comparison results for 'Siemens star'

| | Strategy | | Sampling ratio | | | | | | | |
|---|---|---|---|---|---|---|---|---|---|---|
| | | | 1% | 5% | 10% | 15% | 20% | 40% | 60% | 80% |
| PNSR (dB) | Hadamard | circular | 8.01 | 10.24 | 11.12 | 12.21 | 14.02 | 16.17 | 18.21 | 23.70 |
| | | square | 7.98 | 9.94 | 10.83 | 11.55 | 13.84 | 15.29 | 17.36 | 20.98 |
| | | zig-zag | 7.98 | 10.16 | 11.36 | 12.16 | 13.82 | 16.39 | 19.37 | 24.05 |
| | Fourier | circular | **8.08** | **11.88** | **14.67** | **16.00** | **17.08** | **21.47** | **25.32** | **30.39** |
| | | spiral | 8.07 | 11.84 | 14.46 | 15.93 | 17.07 | 21.28 | 24.91 | 30.11 |
| | | diamond | 8.05 | 11.82 | 14.39 | 15.77 | 16.81 | 20.93 | 24.39 | 29.88 |
| SSIM (%) | Hadamard | circular | 10.0 | 43.1 | 52.4 | 62.2 | 72.2 | 82.1 | 87.0 | 93.2 |
| | | square | 10.3 | 38.4 | 49.2 | 59.3 | 71.3 | 82.5 | 84.9 | 91.4 |
| | | zig-zag | 10.1 | 41.7 | 53.9 | 60.5 | 70.5 | 81.6 | 88.6 | 94.2 |
| | Fourier | circular | **11.1** | 54.6 | 74.9 | 80.1 | 83.9 | 93.8 | 96.5 | 98.6 |
| | | spiral | 11.0 | 54.0 | **75.0** | **81.3** | **84.3** | **93.9** | **97.1** | **98.7** |
| | | diamond | 9.6 | **55.6** | 71.9 | 78.7 | 82.6 | 92.0 | 95.9 | 98.4 |
| Power (%) | Hadamard | circular | 8.9 | 22.9 | 31.2 | 39.5 | 48.3 | 63.8 | 76.6 | 90.6 |
| | | square | 8.7 | 22.4 | 29.9 | 37.3 | 48.1 | 60.9 | 74.2 | 88.3 |
| | | zig-zag | 8.6 | 22.3 | 32.1 | 39.5 | 47.2 | 64.7 | 79.1 | 90.8 |
| | Fourier | circular | **9.4** | 28.1 | **42.8** | **51.3** | **57.7** | **76.2** | **87.1** | **94.8** |
| | | spiral | **9.4** | 28.0 | 42.6 | 51.0 | 57.6 | 75.8 | 86.7 | 94.5 |
| | | diamond | 9.3 | **28.5** | 42.5 | 50.9 | 57.2 | 75.4 | 86.3 | **94.8** |

As the results shown in Fig. 4 and Table 2, FSI with the circular path gives the best reconstruction for the 'Simens star' image. The results by HSI, as they present, are with mosaic artifacts, especially for the oblique features. It is because, Hadamard basis patterns are mosaic look-alike, but lack of oblique features.

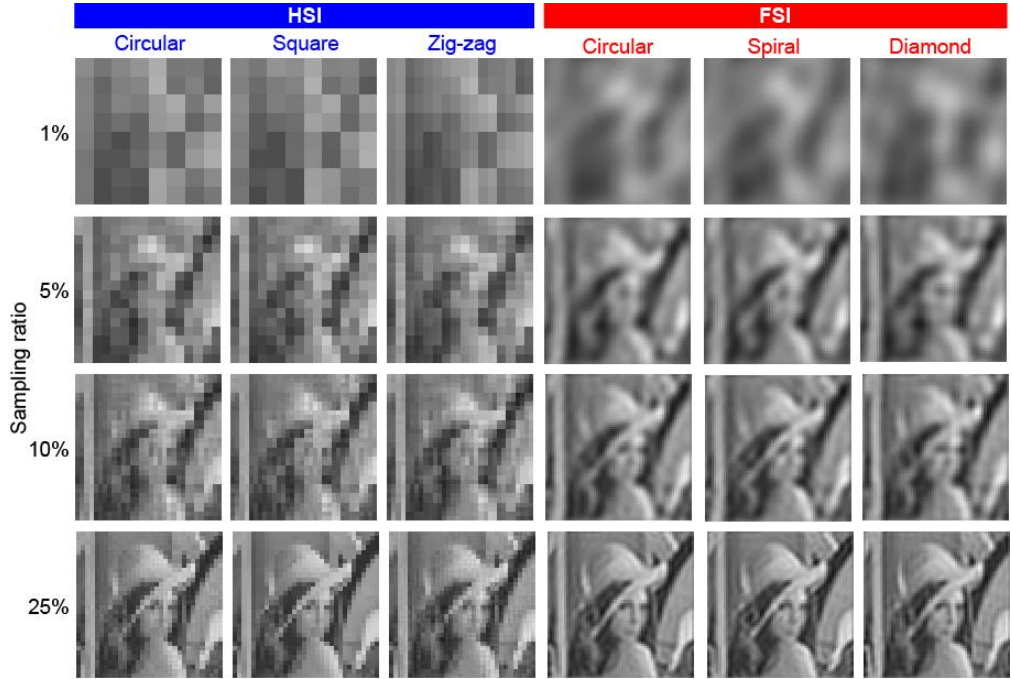

Fig. 5. Comparison results for 'Lena' image reconstruction by HSI and FSI for different sampling ratios.

**Table 3. Quantitative comparison results for 'Lena'**

| | | Strategy | Sampling ratio | | | | | | | |
|---|---|---|---|---|---|---|---|---|---|---|
| | | | 1% | 5% | 10% | 15% | 20% | 40% | 60% | 80% |
| PNSR (dB) | Hadamard | circular | 16.65 | 19.54 | 20.77 | 21.74 | 22.85 | 24.71 | 27.16 | 31.48 |
| | | square | 16.70 | 19.46 | 20.61 | 21.43 | 22.44 | 24.32 | 26.02 | 29.58 |
| | | zig-zag | 16.76 | 19.38 | 20.93 | 21.92 | 22.69 | 25.26 | 28.07 | 32.16 |
| | Fourier | circular | 17.28 | **20.76** | **22.70** | 23.96 | **25.12** | **29.22** | **33.65** | **39.29** |
| | | spiral | **17.35** | 20.70 | 22.50 | 23.96 | 25.07 | 28.90 | 32.99 | 37.93 |
| | | diamond | 17.26 | 20.70 | 22.63 | 23.91 | 24.84 | 29.07 | 33.24 | 38.61 |
| SSIM (%) | Hadamard | circular | 27.1 | 49.6 | 60.2 | 67.3 | 75.7 | 84.2 | 89.7 | 95.8 |
| | | square | 27.0 | 48.9 | 58.9 | 64.9 | 72.4 | 83.1 | 87.9 | 92.9 |
| | | zig-zag | **30.4** | 48.0 | 61.7 | 69.6 | 73.6 | 85.1 | 92.2 | 96.7 |
| | Fourier | circular | 27.3 | 58.4 | **72.6** | **79.6** | **83.8** | 92.4 | 96.9 | **99.2** |
| | | spiral | 27.6 | **58.5** | 71.2 | 79.3 | **83.8** | 92.0 | 96.2 | 98.5 |
| | | diamond | 26.5 | **58.5** | 72.2 | 79.5 | 83.0 | **92.7** | **97.0** | **99.2** |
| Power (%) | Hadamard | circular | 15.6 | 30.2 | 39.3 | 46.6 | 53.2 | 67.3 | 79.3 | 90.8 |
| | | square | 15.8 | 30.4 | 38.7 | 45.4 | 51.9 | 65.6 | 76.8 | 88.7 |
| | | zig-zag | 16.0 | 29.8 | 39.9 | 47.0 | 52.8 | 68.9 | 81.1 | 91.5 |
| | Fourier | circular | 17.6 | 35.3 | 46.3 | 53.6 | **59.8** | **76.8** | **87.8** | **95.3** |
| | | spiral | 18.0 | **35.4** | 46.2 | 53.7 | **59.8** | 76.7 | 87.5 | 94.7 |
| | | diamond | **18.1** | **35.4** | **46.4** | **53.8** | 59.4 | 76.6 | 87.7 | 95.1 |

As the results shown in Fig. 5 and Table 3, FSI outperforms HSI for the 'Lena' image which is a natural image. However, both techniques introduce observable artifacts when sampling ratio is too low. HSI introduces mosaic artifacts while FSI introduces ringing artifacts.

Without loss of generality, we test all images in the USC-SIPI image database [31]. The database consists of 632 images and all the images are categorized into four groups—textures, aerials, miscellaneous, and sequences. Similarly, for every single image, we evaluate reconstruction quality of the six sampling strategies using PSNR, SSIM, and power ratio. The statistical comparison results are derived by counting the overall best for each image. The results are shown in Fig. 6. Based on the comparison results, it is found that FSI has better performance than HSI in terms of reconstruction quality in the situations of undersampling. It is also found that for natural images reconstruction circular path is the best sampling strategy for FSI and zig-zag path is the best sampling strategy for HSI.

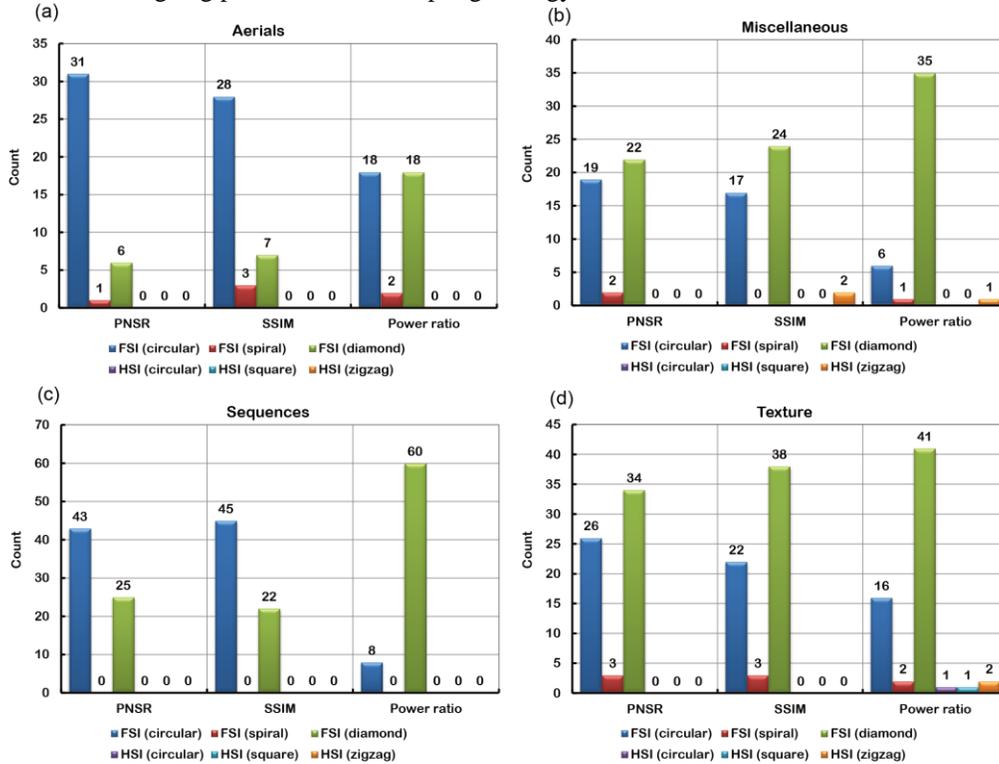

Fig. 6. Statistical comparison results for all four different kinds of images in the USC-SIPI Image Database by HSI and FSI.

These simulations demonstrate that quality of sub-Nyquist sampled images depends on the energy concentration ability of the utilized transformation and the sampling strategy. For natural images, Fourier transform has better energy concentration ability than Hadamard transform, because the correlation between a Fourier basis pattern and a natural image is larger than that between a Hadamard pattern and a natural image. Or in other words, Fourier basis patterns are more similar to the natural images than Hadamard basis patterns. Thus, FSI outperforms HSI in terms of reconstruction quality under sub-Nyquist sampling conditions.

We further compare the both techniques in terms of robustness to noise. To simulate cases of different noise levels, we add white Gaussian noise to the raw data (that is, measurements $D$) resulting in different SNRs. The results are shown in Fig. 7. As the figure shows, HSI is more robust to FSI. It turns out that there's a tradeoff between energy concentration and robustness against noise.

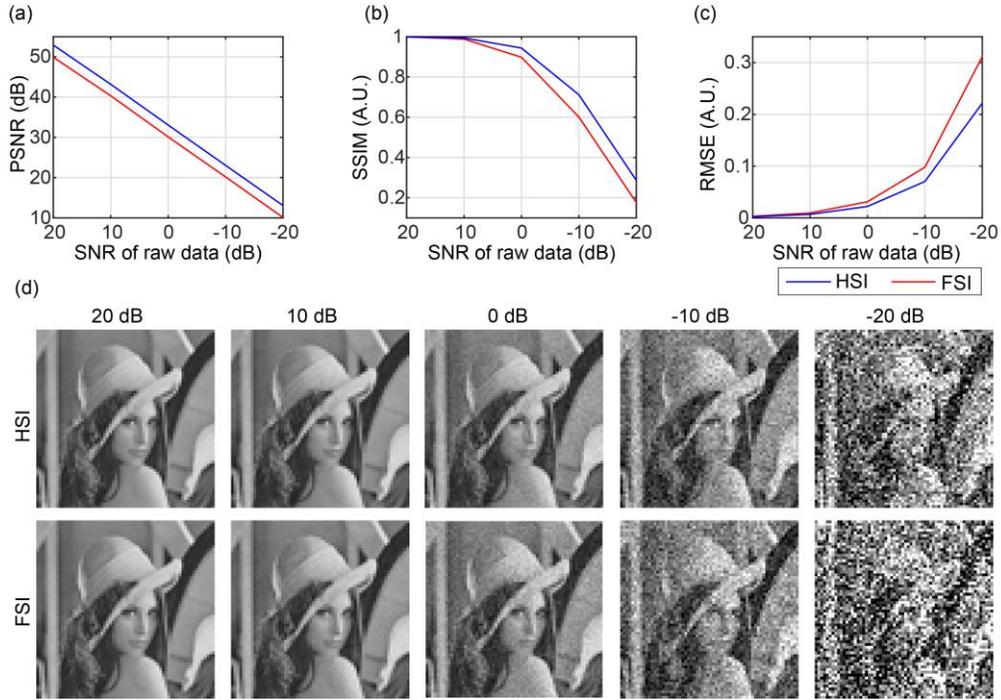

Fig. 7. Noise-robustness comparison. (a) PSNR, (b) SSIM, (c) RMSE, and (d) reconstructed images for different SNRs.

*3.2 Experiments*

We further compare HSI and FSI with experimental data. The first experiment is single-pixel photography using a commercial digital projector (Acer K750) for spatial light illumination. The experimental set-up is shown in Fig. 8. The projector switches pattern every 0.2 seconds. A photodiode (HAMAMATSU S1227-1010BR) is used as a single-pixel detector. The photodiode is driven by a customized amplifier circuit. The resultant electric signals are delivered to a data acquisition board (National Instruments USB-6343 (BNC)). The digitalized data is finally collected by the computer. The methods in comparison are differential HSI and 4-step FSI.

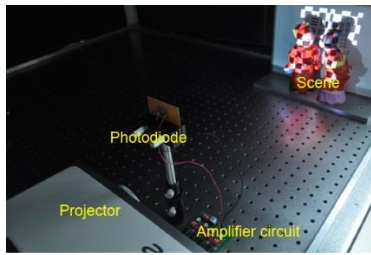

Fig. 8. Experimental set-up for single-pixel photography where a commercial digital projector is used for illumination.

The results are shown in Fig. 9. The resolution of the reconstructed images is 256×256 pixels. As the figure shows, the FSI presents more clear and sharp reconstructions than HSI does in the case of undersampling. The advantage of FSI is relatively observable for the sampling ratio lower than 10%. The results in this experiment coincide with those derived in the numerical simulations.

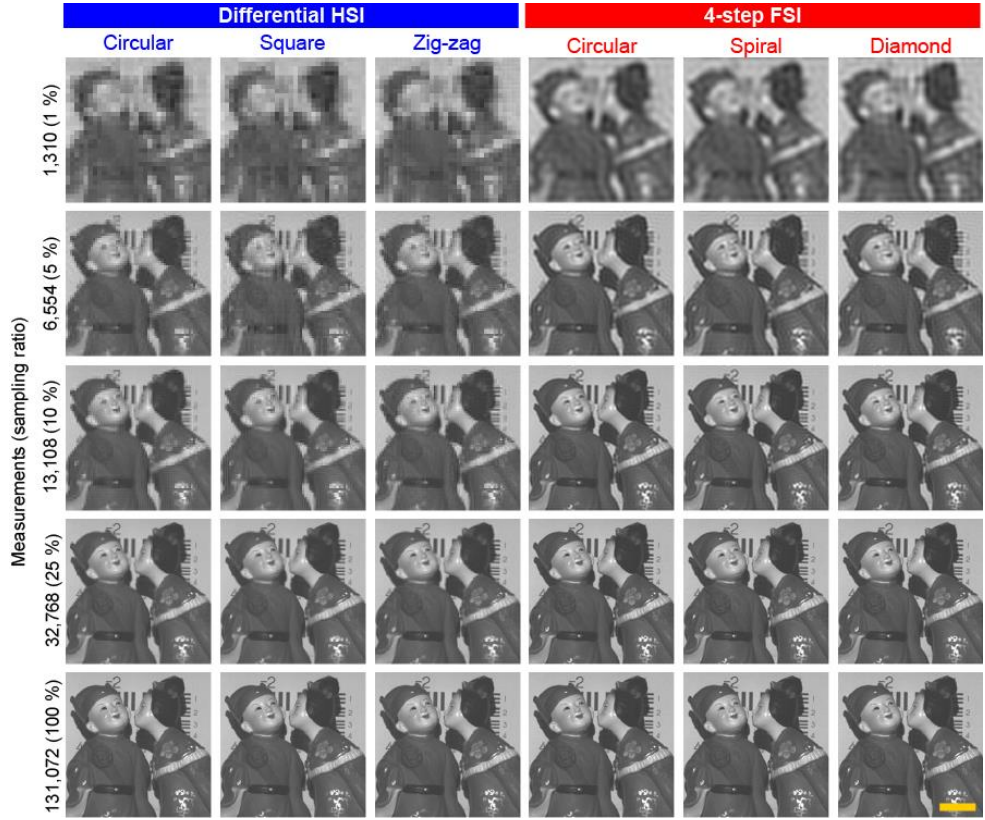

Fig. 9. Comparison results of single-pixel photography. Scale bar = 4 cm.

The second experiment is fast single-pixel imaging. We employ a DLP development kit to achieve high-speed illumination. The DLP development kit is equipped with a 0.7-inch DMD. The DMD has 1024×768 micro mirrors, each of which is 13.6×13.6 μm$^2$ in size. The light source is a 3-watt white LED. The DMD operates at 2,000 Hz, allowing 2,000 binary patterns projection per second. The backscattered light is detected using a photomultipliers tube (Thorlabs PMM01). The resultant electronic signals are transferred to the computer via a data acquisition board [National Instruments USB-6343 (BNC)]. The object to be imaged is some stationery and a piece of A4 paper printed an enlarged 1951 USAF resolution test pattern. The object is under illumination by basis patterns. The methods in comparison are differential HSI 3-step binary FSI, and 4-step binary FSI. The reason why we use binary FSI in this experiment for comparison is that the original FSI is not able to directly use a high-speed DMD for high-speed imaging. It is because the original FSI uses grayscale patterns for illumination, while even an edge-cutting DMD can only display ~250 8-bit grayscale patterns per second, which is a shortcoming of the original FSI.

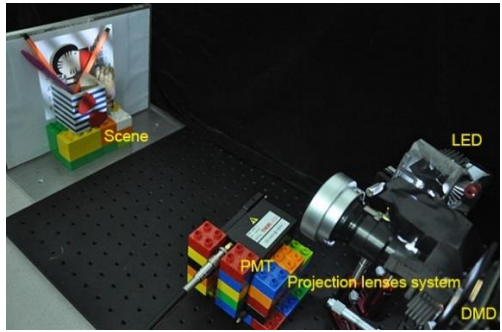

Fig. 10. Experimental set-up for fast single-pixel photography where a DMD development kit is used for illumination.

The resolution of illumination patterns is 256×256 pixels. For binary FSI, the patterns are upsampled using 'bicubic' interpolation so that the resolution of the images becomes 512×512 pixels, twice of the original. Floyd-Steinberg dithering is then applied to the upsampled Fourier basis patterns. For HSI, the Hadamard basis patterns are upsampled to be 512×512 pixels using 'nearest' interpolation. The comparison results are presented in Fig. 11.

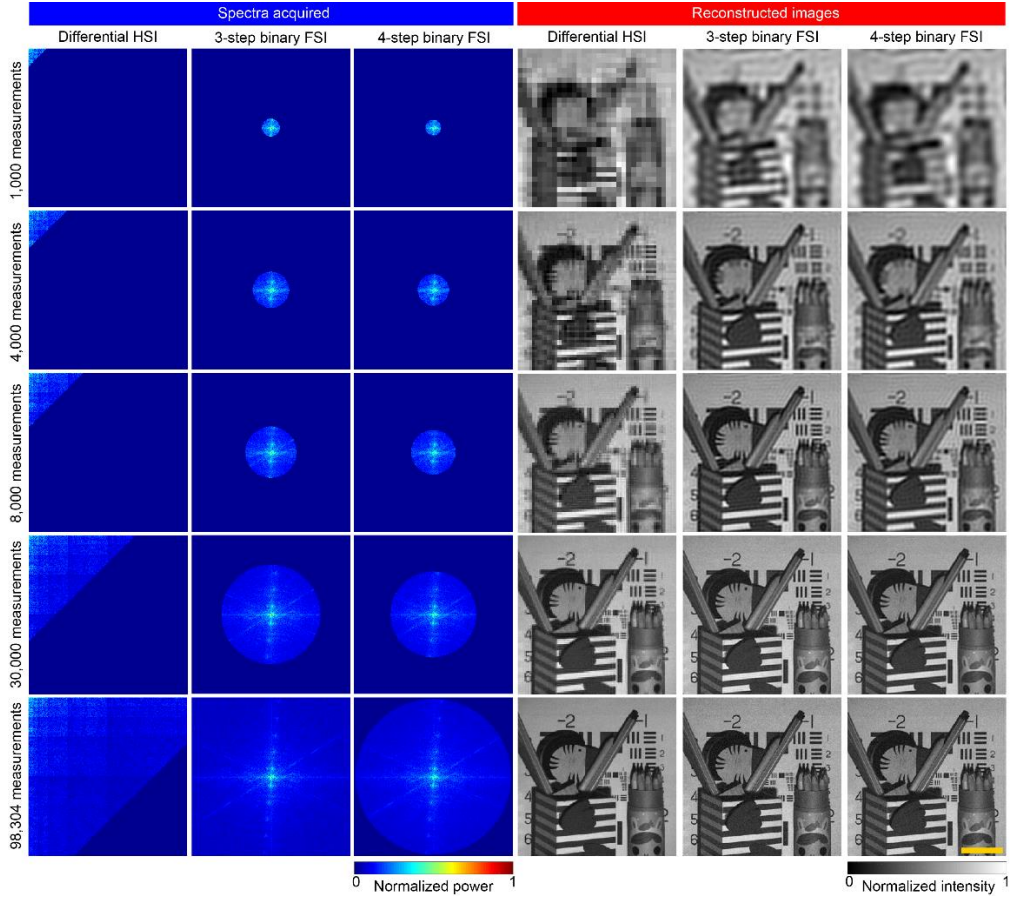

Fig. 11. Comparison results for fast single-pixel photography. Scale bar = 4 cm.

Based on the comparison results, binary FSI outperforms HSI in terms of reconstruction quality in the case of a small number of measurements. In the case of 1,000 measurements, heavy ringing artifacts present in the results of FSI while HSI gives a mosaic reconstruction. 3-step FSI is the only technique that can reconstruct the horizontal lines on the pencil holder, showing its advantage in extremely undersampled cases. In the case of 4,000 measurements, 3-step binary FSI gives better reconstruction, which is evidenced by that all five pencils in the holder on the right become distinguishable. In the case of 8,000, the results by both FSI techniques are satisfactory, except that there still ringing artifacts on the background while the result by HSI is of mosaic, especially for the oblique structures such as the pencil in the left pencil holder. HSI outperforms FSI in terms of uniform pattern reconstruction. The background in the reconstruction by HSI is free of artifacts or noises. In the case of 30,000 measurements, the result by HSI looks as good as those by FSI. In the case of 98,304, the result by HSI is even better than those by FSI in terms of noise level. The noise on the background of the results by FSI becomes noticeable, which is due to the quantization errors.

In short, in the case of a small number of measurements, FSI outperforms HSI for FSI has better energy concentration ability than HSI. In the case of a large number of measurements, HSI has better reconstruction quality than binary FSI. It is because binary FSI introduces quantization errors when generates binary Fourier basis patterns using dithering.

The third experiment is microscopy where we use a digital light projector (Texas Instruments LightCrafter Display 4710 evaluation module) along with a tube lens and an objective lens (Olympus 10X objective (NA = 0.4)) for illumination patterns generation. The experiment set-up is shown in Fig. 12. The DMD of the projector consists of $1920 \times 1080$ micro mirrors whose pitch is 5.4 microns. A Si amplified photodetector (Thorlabs PDA-100A) is used as a single-pixel detector that collects the transmitted light through the object. The model of the data acquisition board used in this experiment is National Instruments USB-6363 (BNC). The object to be imaged is a USAF1951 resolution target (Ready Optics #2015A). Please note that to make full use of the area the DMD, we use 4×4-pixel binning in this experiment. The resolution of illuminations patterns is 256×256 pixels and each pattern uses 1024×1024 mirrors for illumination. The illumination patterns switch every 0.2 seconds.

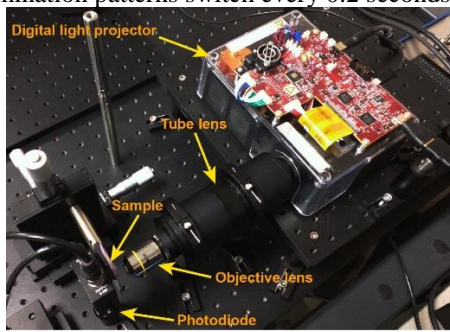

Fig. 12. Comparison results for single-pixel microscopy.

Figures 13 and 14 show the comparison results for single-pixel microscopy. Please note that Fig. 14 shows the partial enlargement of the results in Fig. 13.

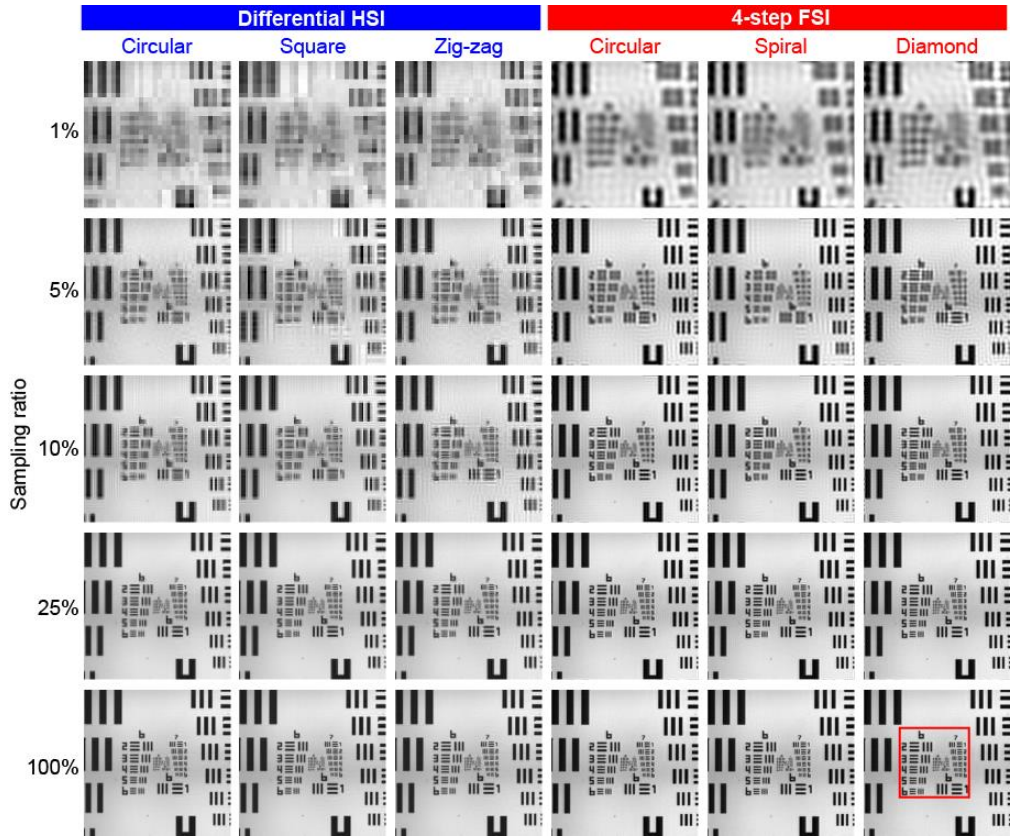

Fig. 13. Comparison results for single-pixel microscopy. The enlarged partial image (in the red box) are shown in Fig. 14.

As the results shown, FSI presents clearer and sharper reconstruction than HSI in the case of undersampling, especially when the sampling ratio is under 10%. For instance, the digits and the bars of the 6[th]. group become distinguishable for the results by FSI. For the full-sampled cases, HSI presents as good results as FSI does.

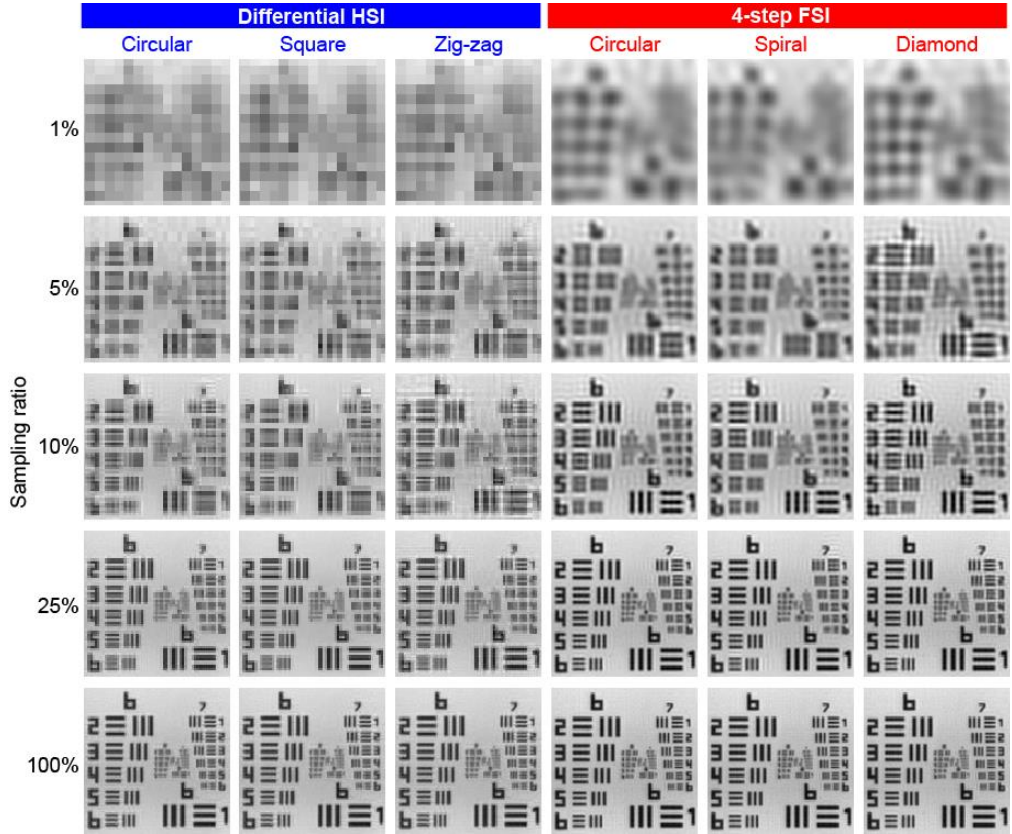

Fig. 14. The partial enlargement of the images shown in Fig. 13.

## 4. Conclusion

We present a systematic comparison between the HSI and the FSI in principle and experiments. The overall comparison is summarized in Table 4. According to the principle analysis and the experimental results, we conclude that the FSI is more efficient than the HSI as Fourier transform better concentrates image energy than Hadamard transform does. Additionally, the 3-step FSI allows measurement reduction in the case of differential measurement. Thus, for applications where efficiency is concerned, the FSI is the primary selection. We also conclude that the HSI has better noise robustness than the FSI and is perfectly suitable for DMD-based single-pixel imaging systems. Therefore, for applications where image quality or accuracy is concerned, the HSI is the primary selection.

**Table 4. Comparison between HSI and FSI**

| | HSI | FSI | |
| --- | --- | --- | --- |
| | | Original FSI | Binary FSI |
| Perfect reconstruction | Yes | Yes | No |
| Measurements for each coefficient (direct sampling) | 1 | 1 | 1 |
| Measurements for each coefficient (differential sampling) | 2 | 1.5 (3-step); 2 (4-step) | 1.5 (3-step); 2 (4-step) |
| Grayscale levels | Binary | Multiple | Binary |
| Robust to dark / read-out noise | Yes | Yes | Yes |
| Robust to quantization errors | Yes | No | Yes |
| Reconstruction for arbitrary-size image | No | Yes | Yes |
| Patterns generation methods | Spatial light modulator | Spatial light modulator / Interference of planar waves | Spatial light modulator |

**Funding.** This work was supported by National Natural Science Foundation of China (NSFC) (61475064). G. Zheng was supported by NSF DBI 1555986, NIH R21EB022378, and NIH R03EB022144.

**Acknowledgments**. The authors thank Dr. Shiping Li for her help with the experimental equipment preparation, and Qinqiu Fang for linguistic assistance.